\documentclass[10pt,conference, table]{IEEEtran}
\pdfoutput=1
\usepackage{etc}
\usepackage[utf8]{inputenc}
\usepackage{graphicx}
\usepackage{cite}
\usepackage{hyperref}
\usepackage{pgfplots}
\usepackage{footmisc}
\usepackage{tabularx}  
\usepackage{booktabs}
\usepackage{colortbl} 
\usepackage{multirow}
\usepackage{balance}
\usepackage{tikz}

\newcommand*\circled[1]{\tikz[baseline=(char.base)]{
            \node[shape=circle,draw,fill=black,text=white,inner sep=2pt] (char) {#1};}}
            
\newcommand{\landfill}{{\sc Landfill}}
\newcommand{\dacos}{{\sc Dacos}}
\newcommand{\dacosx}{{\sc DacosX}}
\newcommand{\tagman}{{\sc Tagman}}

\title{DACOS---A Manually Annotated Dataset of\\Code Smells}

\author{\IEEEauthorblockN{Himesh Nandani, Mootez Saad, Tushar Sharma}
\IEEEauthorblockA{\textit{Dalhousie University} \\
Halifax, Canada \\
\{himesh.nandani, mootez, tushar\}@dal.ca}
}

\date{June 2022}

\begin{document}

\maketitle

\begin{abstract}
Researchers apply machine-learning techniques for code smell detection to counter the subjectivity of many code smells. 
Such approaches need a large, manually annotated  dataset for training and benchmarking.
Existing literature offers a few datasets;
however, they are small in size and, more importantly, do not focus on the subjective code snippets.
In this paper, we present \dacos{}, a manually annotated dataset containing $10,267$ annotations for $5,192$ code snippets. The dataset targets three kinds of code smells at different granularity--\textit{multifaceted abstraction, complex method,} and \textit{long parameter list}. The dataset is created in two phases. The first phase helps us identify the code snippets that are potentially subjective by determining the thresholds of metrics used to detect a smell. The second phase collects annotations for potentially subjective snippets. We also offer an extended dataset \dacosx{} that includes definitely benign and definitely smelly snippets by using the thresholds identified in the first phase.
We have developed \tagman{}, a web application to help annotators view and mark the snippets one-by-one and record the provided annotations. We make the datasets and the web application accessible publicly. This dataset will help researchers working on smell detection techniques to build relevant and context-aware machine-learning models. 
\end{abstract}

\section{Introduction}

Code smells are symptoms of poor design and implementation~\cite{Fowler1999}.
Existing literature shows that code smells have a negative impact on maintainability \cite{Mikhail2011, Gabriele2012},
development effort \cite{Sjoberg2013, Zephyrin2016},
and reliability \cite{Ikama2022, Foutse2012, Tracy2014, Fehmi2013} among other quality attributes. 
Given its importance, the software engineering community has put
significant effort to study various dimensions,
such as their characteristics, impact, causes, and detection mechanisms,
related to code smells \cite{Sharma2018}.

Many code smells are subjective in nature \cite{Sharma2018} \ie{}
a snippet may exhibit a smell in one context, but a similar snippet may not be considered smelly in another context.
\textit{Context} includes the used programming language, experience of the development team, and quality-related practices followed in an organization.
A simple example of smells' subjectivity is a method with, for example, $80$ lines of code.
Based on the context, it could be a \textit{large method} for some developers;
others might not classify the method as a large method.
However, a method with $500$ lines of code will be \textit{definitely} a large method for all developers.

Currently, the majority of commonly used tools use metrics and heuristics to identify code smells \cite{Sharma2018}.
It is often argued that due to the subjective nature of smells,
one cannot come up with universally accepted metric thresholds to classify a snippet in a smelly or benign code in all contexts. \cite{Azadi2018, Sharma2018}
To overcome the challenge introduced by the subjective nature of smells,
researchers propose smell detection using machine-learning techniques \cite{Azadi2018, Pecorelli2019, DiNucci2018, Asif2019, Sharma2021, Sharma2022b}.
Such approaches rely on a code smells dataset, 
ideally manually annotated, 
to train a machine-learning model.
However, existing datasets offer little on multiple fronts.
First, the literature offers only a handful of datasets
such as \landfill{} \cite{Palomba2015b}.
Second, existing datasets contain a small number of annotated samples;
for example, \landfill{} offers annotations for only $243$ snippets.
A dataset with a small number of annotated samples would help a little
to train state-of-the-art deep-learning models with reasonable accuracy.
Next, existing code smells datasets do not filter out code snippets
that are definitely benign or smelly.
For example, a snippet with a very few (say, three)
lines of code cannot have a \textit{long method};
similarly, a snippet with a very large (\eg{} $200$)
number of lines of code definitely suffers from a \textit{long method} smell.
Given that the \textit{value}, in terms of effectiveness,
of a smell dataset lies in the captured subjectivity,
such definite snippets, either definite benign or smelly,
reduce the efficacy offered by a dataset.
Lastly, the available support for different types of smells is limited;
for example, \landfill{} offers annotated snippets for five types of smells.
Given the huge amount of effort involved in annotating code snippets,
the software engineering community needs to complement existing smell datasets for other kinds of actively researched smells.

In this paper, we offer a manually annotated dataset of code smells \textit{viz.} \textbf{\textit{\underline{DA}taset of \underline{CO}de \underline{S}mells}} (\dacos{}).
To create an effective dataset, we filtered the code snippets that are
likely to be subjective by removing the snippets that are either definitely benign or smelly.
This approach helps us better utilize the annotators' effort
by considering their inputs where we actually need them.
The dataset offers annotated code snippets for three code smells---
\textit{multifaceted abstraction} \cite{Suryanarayana2014, Sharma2016b}, \textit{complex method} \cite{Sharma2017}, and \textit{long parameter list} \cite{Fowler1999}.
In addition to a manually annotated dataset on potentially subjective snippets,
we offer \dacosx{} dataset containing a large number of snippets 
that are either definitely benign or smelly.
Furthermore, we developed a web-application \textit{viz.} \tagman{}
to make it easy for annotators
to see one snippet at a time,
and indicate whether a smell is present in the snippet.
We have made source code of \tagman{}\footnote{\url{https://github.com/SMART-Dal/Tagman} \label{github-tagman}}
available publicly.

\begin{figure*}[t]
    \centering
    \includegraphics[width=\textwidth]{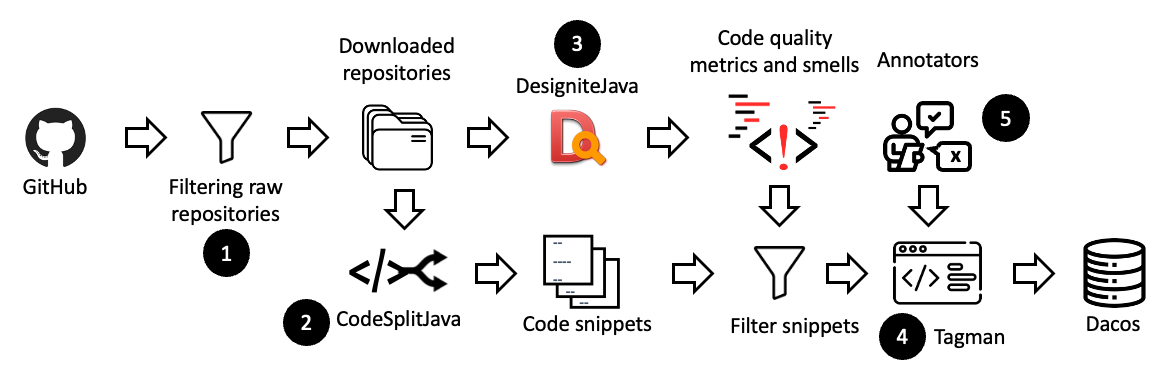}
    \caption{Dataset construction process}
    \label{fig:dataset_construction}
\end{figure*}

We make the following contributions to the state of the art.
\begin{itemize}
    \item We offer \dacos{}, a manually annotated code smell dataset,
    containing $10,267$ annotations for $5,192$ code snippets for the considered code smells.
    
    \item We also provide \dacosx{}, an extended dataset containing $207,605$ snippets that are either definitely benign or smelly. 
    These datasets will help researchers in the field to train and validate their machine-learning models.
    
    \item A configurable web application \tagman{} for easy smell annotations.
    The community may use the application for similar code annotation purposes. 
\end{itemize}

\section{Dataset construction}
Figure \ref{fig:dataset_construction} provides an overview of the dataset construction process.
We elaborate on the steps in detail below.

\subsection{Downloading repositories}
\label{sec:download}

In step \circled{1} from Figure \ref{fig:dataset_construction}, we perform the following tasks to identify and download repositories.

\begin{itemize}

    \item We use \texttt{searchgithubrepo}~\cite{Sharma2022} python package,
    which in turn uses the {\sc GitHub} {\sc GraphQL api} \cite{graphql} to filter {\sc GitHub} repositories. 
    \item To identify high quality Java repositories,
    we select repositories with more than or equal to $13$ thousand stars and more than ten thousand lines of code.
    \item Also, we discard the repositories that are not modified in the last one year. 
    \item 
    In addition, we use QScored \cite{Sharma2021b} to filter out repositories based on their code quality score.
    QScored assigns a weighted quality score based on the detected smells at various granularities. 
    We select repositories that have a quality score less than ten (the higher the score, the poorer the quality).
    \item
    Finally, we obtained ten repositories after applying the filtering criteria.
    We download the selected repositories.
\end{itemize}

\subsection{Dividing the repositories into classes and methods}
We need to split a repository into individual methods and classes so that
\tagman{} can show individual snippets one by one to an annotator.
We use {\sc CodeSplitJava}~\cite{CodeSplitJava}
in step \circled{2}
to split each repository into individual methods and classes.

\subsection{Analyzing repositories}
\label{sec:analyze}
In step \circled{3}, we employ a metrics-based filtering process in the phase-2
of manual annotation.
We use {\sc DesigniteJava} \cite{Sharma2016} to compute code quality metrics.
{\sc DesigniteJava}
computes a variety of code quality metrics and detects smells;
it has been used
in various studies~\cite{Oizumi2019, Sharma2020, Eposhi2019, Uchoa2020, Alenezi2018}.
We elaborate the process to filter out non-subjective samples in the manual annotation step.

\subsection{Tagman}
\begin{figure}[h]
    \centering
    \includegraphics[width=\columnwidth]{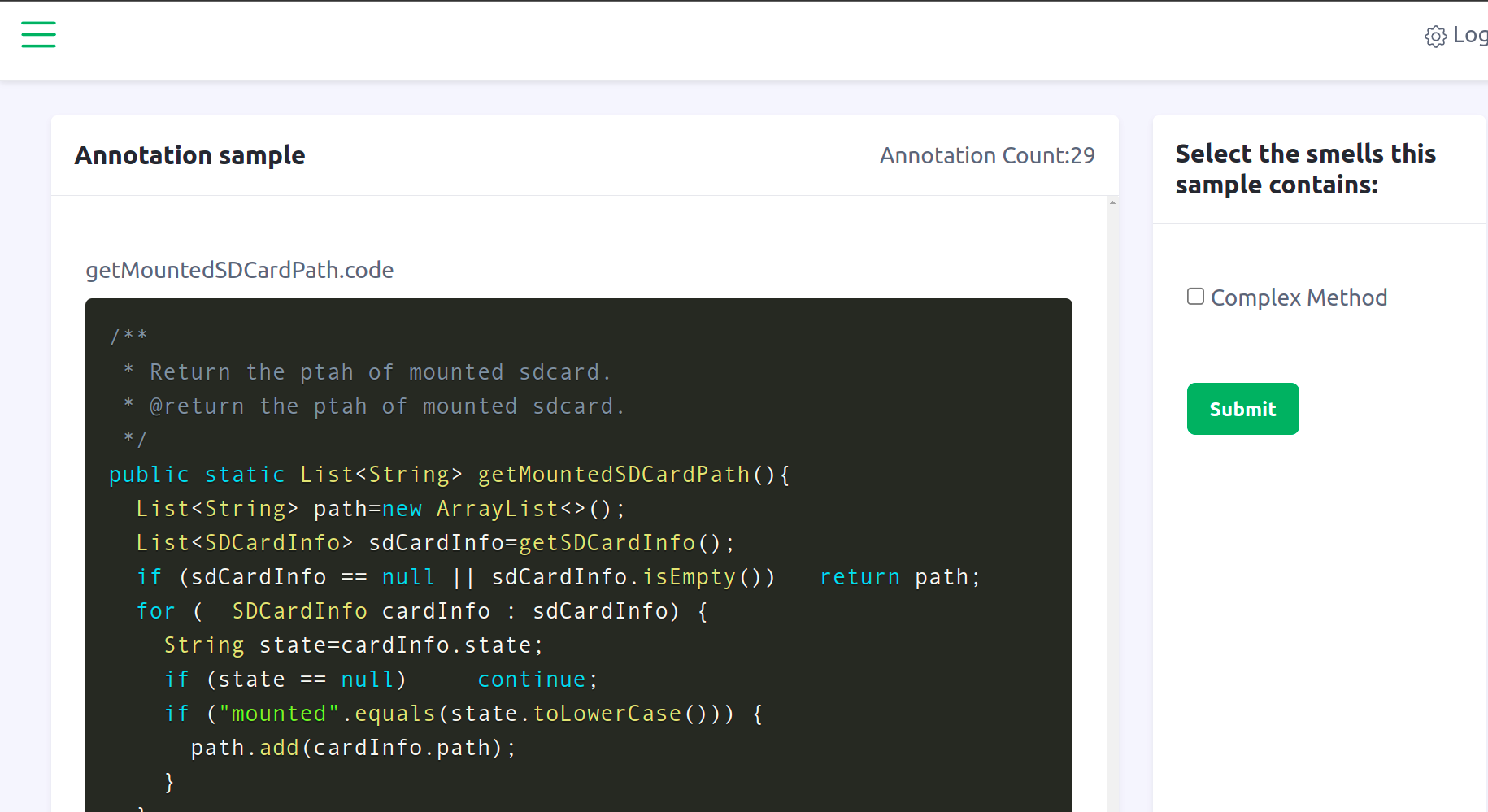}
    \caption{Annotation user interface of \tagman{}}
    \label{fig:tagman}
\end{figure}

\begin{figure}[h]
    \centering
    \includegraphics[width=\columnwidth]{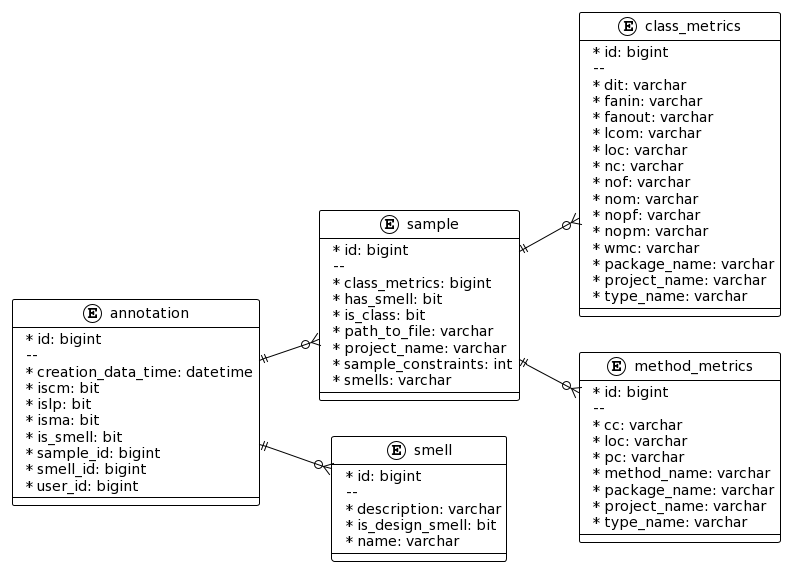}
    \caption{Schema of the \dacos{} database}
    \label{fig:database}
\end{figure}

\tagman{}, shown as \circled{4} in Figure~\ref{fig:dataset_construction},
is a web-based tool that we created to facilitate the annotation process.
Figure \ref{fig:tagman} shows a screenshot of the application showing a code snippet and an option to annotate the snippet with a smell.
The front-end of \tagman{} is written in \textit{Thymeleaf} and {\sc html/css}. 
The back-end of the application is developed in \textit{SpringBootJava} and the data is stored in a {\sc MySQL} database.
Figure \ref{fig:database} shows the schema of the database.

At the beginning of the code smell annotation cycle,
we upload a {\sc csv} file containing the repository names and {\sc url} of
selected {\sc GitHub} repositories. 
\tagman{} back-end uses a set of Python scripts\footnote{\url{https://github.com/SMART-Dal/Tagman-scripts} \label{github-scripts}} to download {\sc GitHub} repositories,
split the code into class and method files,
and run \designitejava{}. 


Once the data import is completed, the tool is ready to start accepting
annotations.
To start an annotation session,
a user login (or sign up) to the application.
Then, the user is presented with
instructions including the definitions of code smells. 
The user can then start annotating the presented code snippets.

\subsection{Manual annotation}
We employ a snippet selection mechanism to identify potentially subjective snippets
\textit{w.r.t.} a code smell.
We do so to improve the effectiveness of the resultant dataset by only including manual annotations for potentially subjective code snippets.
Also, such a strategy helps us better utilize the available annotators' time.
A potentially subjective snippet is a code snippet that may get classified as benign or smelly based on context.
The rest of the snippets that are not identified as potentially subjective snippets
are either definitely benign or smelly snippet.
For example, \textit{cyclomatic complexity} ({\sc cc})  \cite{McCabe1976} is commonly used to detect \textit{complex method} smell.
A code snippet is definitely benign if {\sc cc} is very low (\eg{} {\sc cc}=1);
similarly, a snippet is definitely smelly if {\sc cc} is very high for a method (\eg{} {\sc cc}=$30$).
We divide our annotation process into two phases.
In the first phase, we show all snippets, \ie{} without any filtering,
to annotators to identify metrics thresholds to determine whether a snippet is
potentially subjective or not.
The second phase uses the identified metrics thresholds and show the
filtered code snippets to
annotators.





\vspace{2mm}
\subsubsection{Phase-1}
In the first phase, we show code snippets without any filtering to annotators.
\tagman{} presents one snippet at a time to the annotators and collects their response on whether the shown snippet has a code smell or not.
We show a code snippet to two randomly chosen annotators and record their responses.
The annotators recruited, on a volunteer basis,
for this phase were graduate students of Computer Science
enrolled in a software engineering course (during summer 2022)
that cover code smells extensively.
A total of $110$ annotators participated in this phase.




We compute the minimum and maximum threshold for metrics that are used to decide
the presence of a code smell based on the collected responses in Phase-1.
We received a total $17,869$ responses in this phase.
We compute the lowest metric value (t$_l$) where the smell is identified,
for each smell individually,
to obtain the threshold at the lower side.
Similarly, we extract the highest metric value (t$_h$) where the smell is \textit{not} identified.
Then, we compute the standard deviation (\texttt{sd}) of the metric value for the samples where the smell is identified.
Finally, we obtain the low threshold using \texttt{max(m$_l$, t$_l$ - sd)}
and high threshold using \texttt{min(m$_h$, t$_h$ + sd)} for subjective snippet identification.
Here, m$_l$ and m$_h$ represent the lowest and highest possible values of a metric.
Table \ref{table:phase1} summarizes the quality metrics corresponding to each
code smell and their thresholds for identifying
subjective snippets.
For instance,
for \textit{cyclomatic complexity} metric, we obtain $4$ and $7$ from the above calculation after rounding the values to the nearest integer.

\begin{table}[h!]
\rowcolors{2}{gray!25}{white}
\caption{Code quality metrics used for code smells and their low and high thresholds for subjective snippet identification}

\begin{tabular}{p{2.2cm}%
   p{3cm}
   >{\raggedleft\arraybackslash}p{2.2cm}
  }
\textbf{Code smell}    & \textbf{Code quality metric} & \textbf{Metric threshold} \\
  \toprule{}
Complex method 
& Cyclomatic complexity 
&   4--7 \\
Long parameter list      
& Parameter count                              
&     3--6          \\
Multifaceted abstraction 
& Lack of cohesion among methods (LCOM)        
&   0.4--0.8 \\\bottomrule

\end{tabular}
\label{table:phase1}
\end{table}

\vspace{2mm}
\subsubsection{Phase-2}
We configure our filtering mechanism based on the thresholds obtained from Phase-1
and invite annotators by advertising the link of \tagman{} installation
on social media platforms such as Twitter and LinkedIn.
The invitation was open to all software developers, software engineering students,
and researchers who understand Java programming language and at least basic object-oriented concepts.
We kept the invitation open for six weeks during Dec-Jan 2022-23.
A total of $82$ annotators participated in this phase.
\tagman{} showed snippets that have metric values falling between the low and high
thresholds (inclusive).
We configured \tagman{} to get two annotations for each sample to improve the reliability of the annotations.

\subsection{Dataset information}
After the phase-2, we received a total of $10,267$ annotations for $5,192$
samples from $86$ annotators.
Table \ref{table:dataset-metadata} presents the number of annotations and samples per
smell type.

\begin{table}[h!]
\rowcolors{2}{gray!25}{white}
\caption{Dataset metadata}
\centering
\begin{tabular}{p{1.3cm}p{2.8cm}%
   >{\raggedleft}p{1.5cm}
   >{\raggedleft\arraybackslash}p{1.5cm}
  }
\textbf{Dataset}
&\textbf{Code smell}    & \textbf{\#Annotations} & \textbf{\#Samples} \\
  \toprule
&Complex method 
& 4,349 
&  2,197 \\ 
\cellcolor{white}&Long parameter list      
& 3,221                              
&   1,634        \\
\multirow{-3}{*}{\dacos{}}&Multifaceted abstraction 
& 2,697        
&  1,361  \\\midrule
&\textbf{Total}
& \textbf{10,267}
& \textbf{5,192}\\\midrule
&Complex method 
& --
&  94,489 \\
\cellcolor{white}&Long parameter list      
& --                              
&   93,442         \\
\multirow{-3}{*}{\dacosx{}}&Multifaceted abstraction 
& --     
&   19,674 \\\midrule
&\textbf{Total}
& \textbf{--}
& \textbf{207,605}\\
\bottomrule

\end{tabular}
\label{table:dataset-metadata}
\end{table}

\noindent
\textit{Dataset availability:}
The datasets offered in this paper are available online \cite{Nandani2023}. 
Also, the repositories containing scripts\footref{github-scripts} used to prepare code snippets as well as
code annotation application\footnote{\url{https://github.com/SMART-Dal/Tagman} \label{github-tagman}} \ie{} \tagman{} are available online.

\section{Potential research applications}
\begin{itemize}
    \item 
    \textbf{Detecting and validating code smells:}
Given the subjective nature, traditional smell detection tools
that implement rules and heuristics to identify smells,
do poorly.
The success of a machine-learning approach to detect  smells depends on
the availability of a large manually annotated dataset.
The presented datasets, \dacos{} and \dacosx{}, complement existing datasets
by offering a large number of subjective code snippets for three code smells that
the existing datasets do not cover.

\item
\textbf{Correlating code smells with software engineering aspects:}
A variety of exploratory and empirical studies investigating the impact of code smells
exists.
It includes bug prediction \cite{Palomba2016},
maintainability prediction \cite{Yamashita2012}, and
maintenance effort \cite{Yamashita2012b}.
In addition to existing directions,
the tools trained or fine-tuned from the offered datasets
can be used to effectively establish a relationship between code smells and
productivity of a software development team.

\item
\textbf{Extending \tagman{} for code annotation:}
The code annotation application that we developed to create this dataset can be extended for similar kinds of code annotation, for example, to spot vulnerable code and to segregate well-written identifiers.
\end{itemize}

\vspace{2mm}
\section{Related datasets}


Software engineering literature offers a small number of manually annotated datasets for code smells.
Palomba \etal{} \cite{Palomba2015a}
offered a dataset \landfill{} containing annotations for
five types of code smells---\textit{divergent change,
shotgun surgery, parallel inheritance, blob,} and \textit{feature envy}.
They offered annotations for $243$ snippets.
They also developed an online portal where contributors can annotate code for smells, 
however as of writing this paper (Jan 2023), the portal is not accessible. 
Madeyski \etal{}~\cite{Madeyski2020} proposed {\sc mlcq}---
a code smell manually annotated dataset.
The dataset contains $14.7$ thousand annotations for $4,770$ samples.
The dataset considered four smells---\textit{blob, data class, long method,} and \textit{feature envy}.
Both of the datasets mentioned above do not consider the subjectiveness of a code snippet;
hence most of the snippets might not add any new information for a machine-learning
classifier when used in training.
Also, we chose the code smells that are not covered by any existing code smell dataset and hence complement the existing datasets.
There are some code smells datasets such as QScored\cite{Sharma2021b}.
Though the QScored dataset is large, the samples are not manually annotated and hence lack the required capturing of context.



\section{Threats to validity}

\textit{Internal validity} threats concern the ability to draw conclusions from our experimental results.
In phase-2 of manual annotation,
we invited volunteers with at least a basic understanding of Java programming language and object-orientation concepts. 
We advertised the invitation on social media professional channels (Twitter and LinkedIn).
Given the anonymity of the exercise, we do not have any mechanism to verify the assumption that the participants has sufficient knowledge to attempt the annotations.
However, we offered all the major participants (with at least 50 annotations) to include them as contributors to the dataset;
we perceive such a measure would have motivated the annotators to perform the annotations to the best of their abilities.
Additionally, we configured \tagman{} to obtain two annotations per sample so that we can reduce the likelihood of a random annotation.

\textit{External threats} are concerned with the ability to generalize our results.
The proposed dataset is for snippets written in Java.
However, our code annotation tool is generic and it can be used to annotate snippets from any programming language.
Additionally, scripts used to generate individual snippets can be customized to use any other external tool for splitting the code into methods and classes.
Furthermore, the thresholds used in the annotation process to filter snippets based on low and high thresholds of a metric are configurable.


\vspace{2mm}
\section{Limitations, conclusions, and future work}
We offer \dacos{}---a manually annotated code smell dataset
containing $10,267$ annotations for $5,192$ subjective code snippets.
We also providea large \dacosx{} dataset containing definitely benign and
definitely smelly snippets in addition to those present in \dacos{}.
The paper offers \tagman{}, a code annotation application, that can be reused in similar contexts.

The proposed dataset covers three code smells.
We selected a rather small set of code smells to consider in the dataset because
it is better to have more number of annotations for a smell rather than having small number of annotations per smell. Also, we chose a set of smells that are not covered by existing code smells dataset.
We configured \tagman{} to obtain two annotations per sample.
Though it improves the reliability of the dataset, one may argue that it may introduce a situation where the annotations are contradictory to each other.
We can mitigate the limitation by increasing the number of annotations per sample to three; we will incorporate this mechanism in the future version of the datasets.
Additionally, we would like to expand the scope of the dataset in terms of programming language, number of samples, and number of supported smells in the future.


\balance
\bibliographystyle{IEEEtran}      
\bibliography{references}   

\end{document}